# Breaking molecular nitrogen under mild conditions with an atomically clean lanthanide surface

Felicia Ullstad*, Gabriel Bioletti, Jay Chan, Audrey Proust, Charlotte Bodin, Ben J. Ruck, H. Joe Trodahl, Franck Natali*

MacDiarmid Institute for Advanced Materials and Nanotechnology, School of Chemical and Physical Sciences, Victoria University of Wellington, PO Box 600, Wellington, New Zealand.

**ABSTRACT:** A route to break molecular nitrogen ($N_2$) under mild conditions is demonstrated by $N_2$ gas cracking on, and incorporation into, lanthanide films. Successful growth of lanthanide nitride thin films, made by evaporation of lanthanides in a partial $N_2$ atmosphere at room temperature and pressure as low as $10^{-4}$ Torr, is confirmed using X-ray diffraction. *In-situ* conductance measurements of pure lanthanide thin films exposed to $N_2$ gas show an immediate surface reaction and a slower bulk reaction. Finally, we report partial reversal of the nitrogen incorporation in a lanthanide nitride by cycling vacuum and nitrogen conditions in the sample chamber.

Ammonia ($NH_3$) is among the most important chemicals in today's economy, an irreplaceable precursor in the fertiliser production supporting the world's population.[1] However, the industrial synthesis of $NH_3$ from molecular nitrogen (**$N_2$**) and hydrogen ($H_2$), the Haber–Bosch process, is one of the most severe processes in the chemical industry.[2] The cleavage of the **$N_2$** bond, the main challenge in the industrial $NH_3$ synthesis, is only feasible under extreme conditions of high temperatures and pressures, consuming a few percent of the world's energy production.[2] As a consequence, breaking the immensely strong triple bond of molecular nitrogen has attracted an enormous research effort. Catalysts that could allow a facile breaking of $N_2$, and a potential energy-efficient $NH_3$ synthesis, include typically enzymatic and organometallic approaches as well as electro- and photocatalytic materials, with a strong interest in mimicking biological systems such as nitrogen fixation, as well.[3-13] The focus has concentrated on these relatively complex approaches suffering from a number of drawbacks, e.g. large atomic cluster-based catalysts that do not lend themselves to theoretical treatment and the lack of *in-situ* and real-time characterisation on an atomic level. In stark contrast, here we report on a radically different approach, the facility to use an atomically clean surface of lanthanide (L) metal to break molecular nitrogen at room temperature and under pressure much lower than one atmosphere, typically 7 to 8 orders of magnitude smaller, through both *in-situ* and *ex-situ* measurements.

The primary evidence points for the $N_2$ reacting efficiently with the lanthanide presented in this paper are fourfold. Firstly, we show the formation of lanthanide nitrides (LN) during lanthanide evaporation in a partial nitrogen atmosphere at room temperature. Secondly, we demonstrate a change in electrical conductivity in LN thin films, samarium nitride (SmN) and gadolinium nitride (GdN), when the L:N ratio during deposition is changed. Thirdly, the electrical conductance and crystallographic properties are significantly altered in lanthanide thin films when exposed to $N_2$. Fourth and lastly, the conductance in SmN films when cycling $N_2$ pressure and ultra-high vacuum conditions shows the possibility of removing nitrogen from the SmN lattice.

The deposition and the subsequent N2 exposure of the lanthanide (L) thin layers are carried out in an ultra-high vacuum system, with a base pressure of < $10^{-8}$ Torr. The purity of the as-received L charges is typically 3N or 4N, and the purity of $N_2$ is at least 4N and introduced through a Ni sponge. Several L elements have been evaporated on the surface of an amorphous substrate (fused silica) in the presence of $N_2$ (Figure 1). The substrate is kept at ambient temperature (~30°C), the partial pressure of $N_2$ is ~$10^{-4}$ Torr, and the deposition rate is typically 200-300nm/h, resulting in thin films of thickness a few hundred nm. Surprisingly, most lanthanide mononitrides are successfully formed by simply depositing L elements in the presence of $N_2$ as displayed in Figure 1 where the X-ray diffraction (XRD) patterns are shown. The lanthanides that are most rigidly trivalent (Sm, Gd, Dy, Er, Tb)[14] react with the $N_2$ and form a rocksalt-structured mononitride ($L^{3+}$-$N^{3-}$), showing characteristic (111)-dominated polycrystalline patterns as expected for room temperature grown LN thin layers.[15] As will be illustrated below it is necessary that the $N_2$ flux on the substrate is two or more orders of magnitude larger than the flux of lanthanide atoms to form a stoichiometric LN film, clearly indicating a significant reaction coefficient for $N_2$. As a contrast, the more readily divalent lanthanides (Eu, Yb)[14] do not react with $N_2$ to form a mononitride, but instead form a pure lanthanide thin film. It is possible to grow the divalent lanthanide nitrides, but this requires activated nitrogen.[15] The results seem to suggest that the valency of the lanthanides influences their reaction with $N_2$, and it is also interesting to point out that non-reacting L elements, Eu and Yb, possess the largest atomic volume.[16]

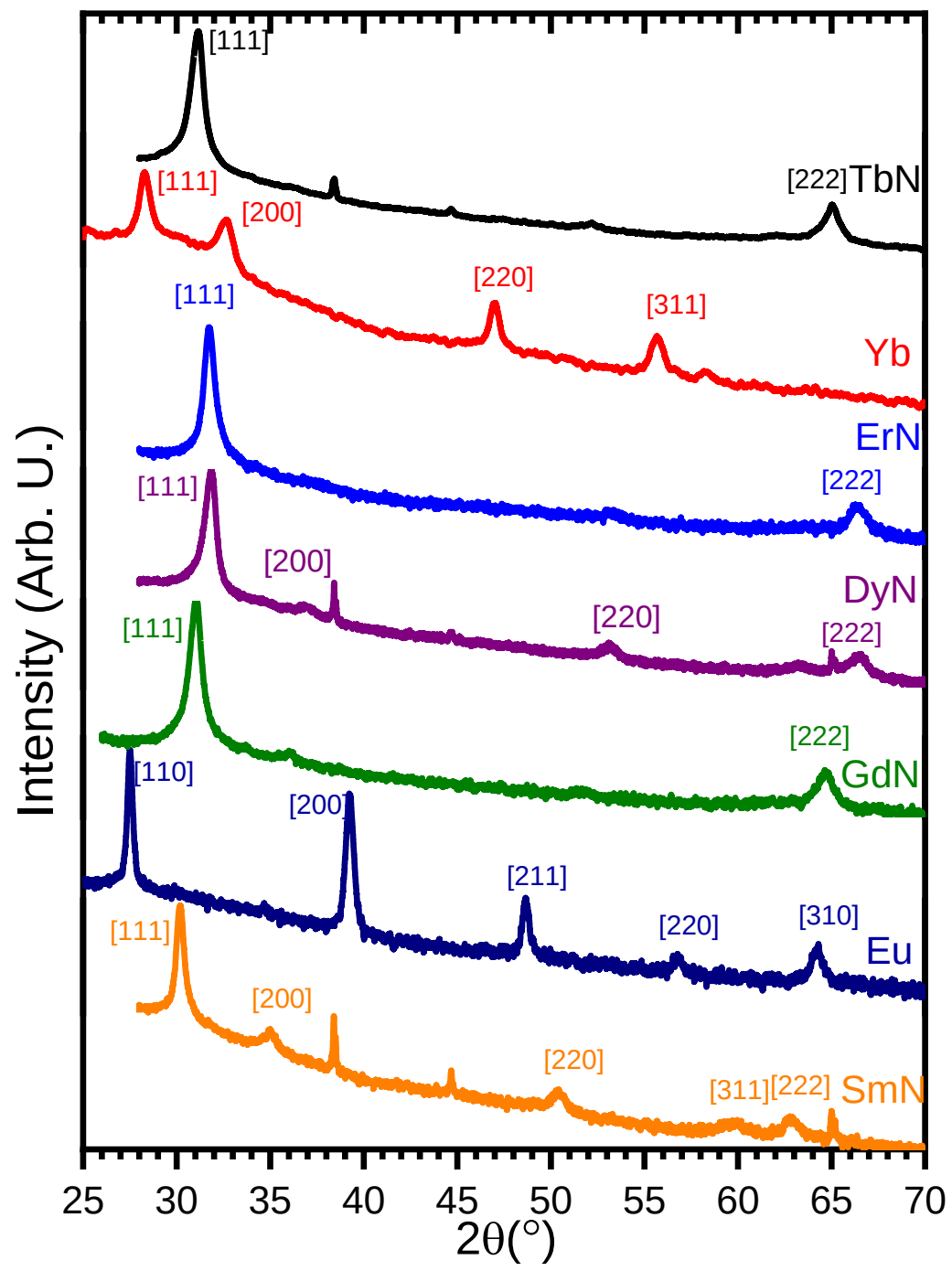

Figure 1. XRD 2θ scans of thin films of lanthanides deposited in a $N_2$ atmosphere. Trivalent lanthanides form a rocksalt-structured mononitride, whereas divalent lanthanides do not react with $N_2$.

To further investigate the reaction coefficient during the L:N2 reaction we perform *in-situ* electric conductance of a GdN and SmN thin films for varying $N_2$ pressures (and hence different L:N ratios) during growth (Figure 2). The films are grown between pre-deposited electric contacts on a SiO2/Si substrate and the electrical resistance is recorded as the films grow. As the nitrogen pressure ($P_N$) in the chamber is lowered, the conductivity of the LN film increases. This increase is expected as lower nitrogen pressure introduces more nitrogen vacancies ($V_N$), acting as n-type dopants in the lattice which donate one to three electrons each.[17,18] The conductivity, supported by Hall measurements, signal a carrier concentration of about 0.1-2 x $10^{21}$ $cm^{-3}$ in films grown in $10^{-4}$ Torr $N_2$. The $V_N$ concentration in LN films is only of the order of 1%, too small to establish accurately with the compositional, i.e. stochiometry, measurements available. At lower $P_N$ the carrier concentration rises with an inverse proportionality to $P_N$, as expected for an $N_2$ sticking coefficient that is proportional to the fraction of surface L ions with unsatisfied nitrogen bonds, which is then in turn proportional to $V_N$.

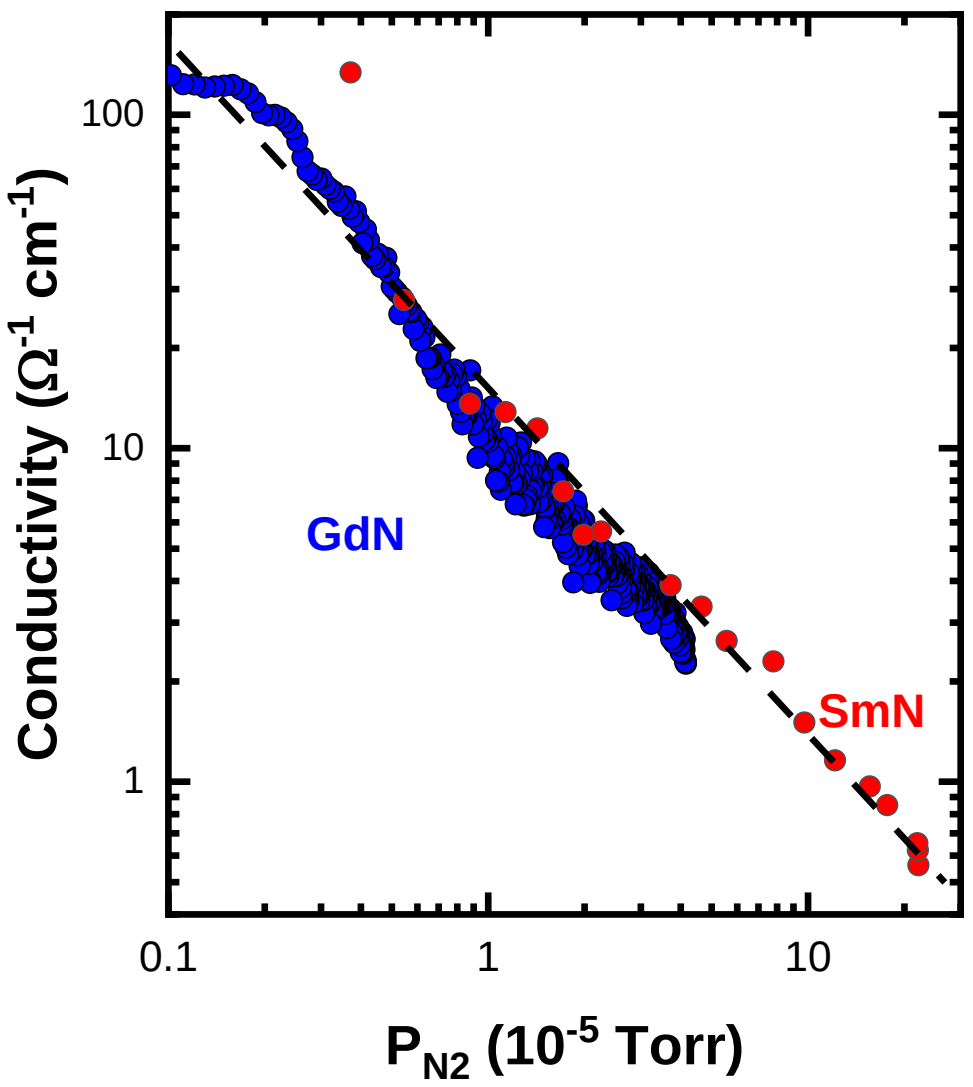

Figure 2. Electrical conductivity of GdN and SmN thin films as a function of the $N_2$ pressure during lanthanide evaporation. GdN data from Ref. 18 (Granville *et al.*). Note that the conductivity of pure Sm and Gd is four orders of magnitude larger.

Our ability to grow LN thin layers at different $N_2$ pressures suggests a catalytic reaction at the surface that breaks the $N_2$ bond to allow the L+N→LN reaction to take place at room temperature and pressure as low as $10^{-4}$ Torr. To glean some information about this reaction we grew a thin (25 nm) film of pure Gd and then exposed it to N2 within the same UHV system. The exposure was carried out at room temperature for 10 minutes and under a partial $N_2$ pressure of 2 × $10^{-4}$ Torr. To confirm the Gd+$N_2$ → GdN reaction, XRD measurements are performed ex-situ after removing the $N_2$-exposed Gd layer from the UHV system. In order to prevent modification in air, the $N_2$-exposed Gd layer was capped with a passivation layer. The Figure 3 shows X-ray diffraction 2θ scan for a 25 nm thick Gd layer after $N_2$ exposure (black). We can see only a peak associated to GdN (111). No trace of pure Gd is detected. For a comparison, a XRD 2theta-omega scan of a pure 25 nm thick Gd lanthanide layer is shown (grey).

We have also monitored the in-situ conductance of some pure lanthanide films (Sm, Gd, Dy) as they were exposed to $N_2$ (Figure 4). The data show the conductance of the films relative to the pure lanthanide films before N2 exposure. The dose is the total $N_2$ gas dose that the films are exposed to in the chamber. It is worth noting that there are over four order of magnitude difference in conductivity between the lanthanide and its lanthanide nitride. This difference in magnitude means that if we view each unit layer of the lanthanide (nitride) thin film as a parallel resistor, we can set the conductance contribution of the lanthanide nitride in such a network to zero.

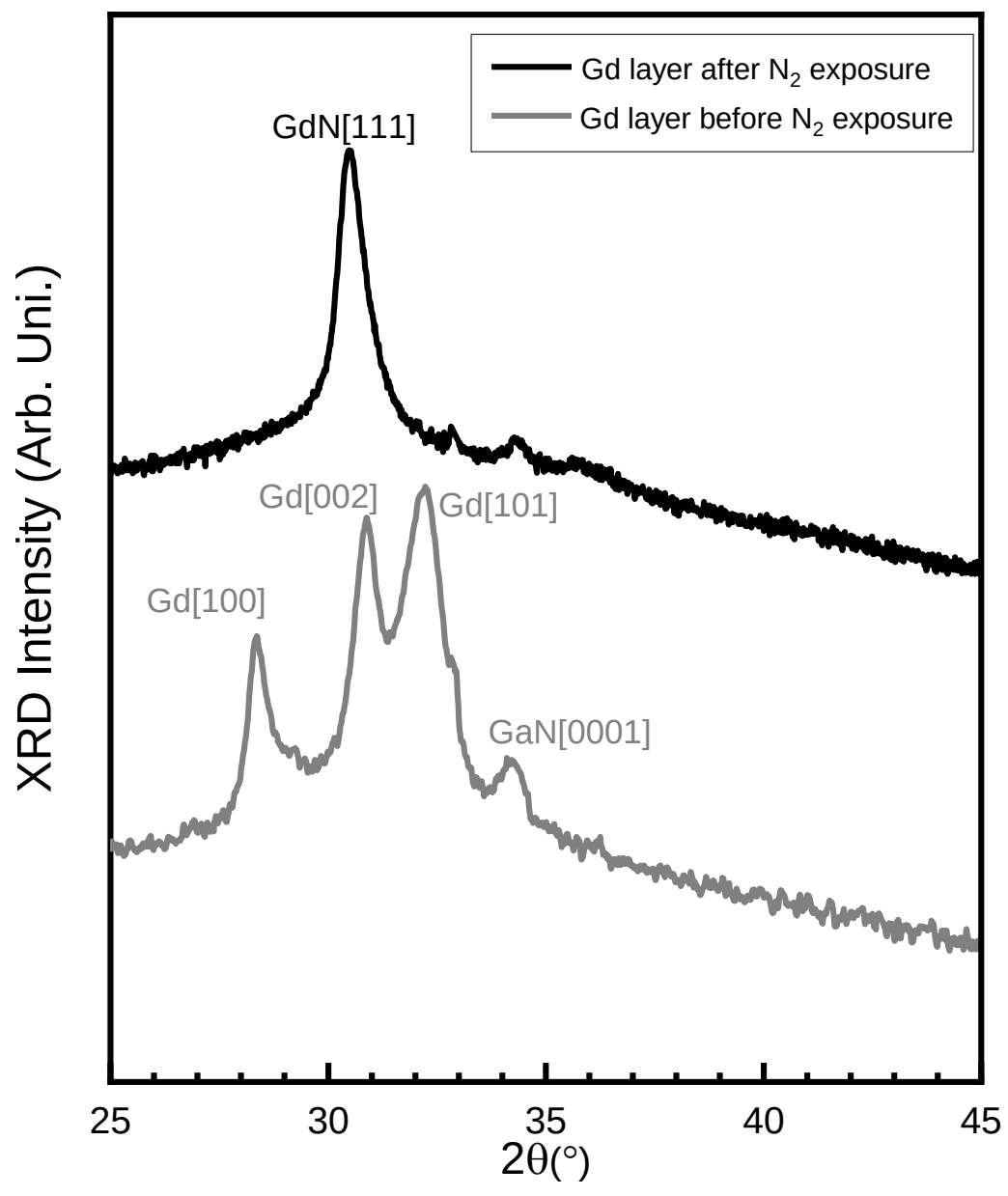

Figure 3. XRD 2θ scans of a 25 nm thick Gd layer after $N_2$ exposure (black) and a pure 25 nm thick Gd lanthanide layer (grey).

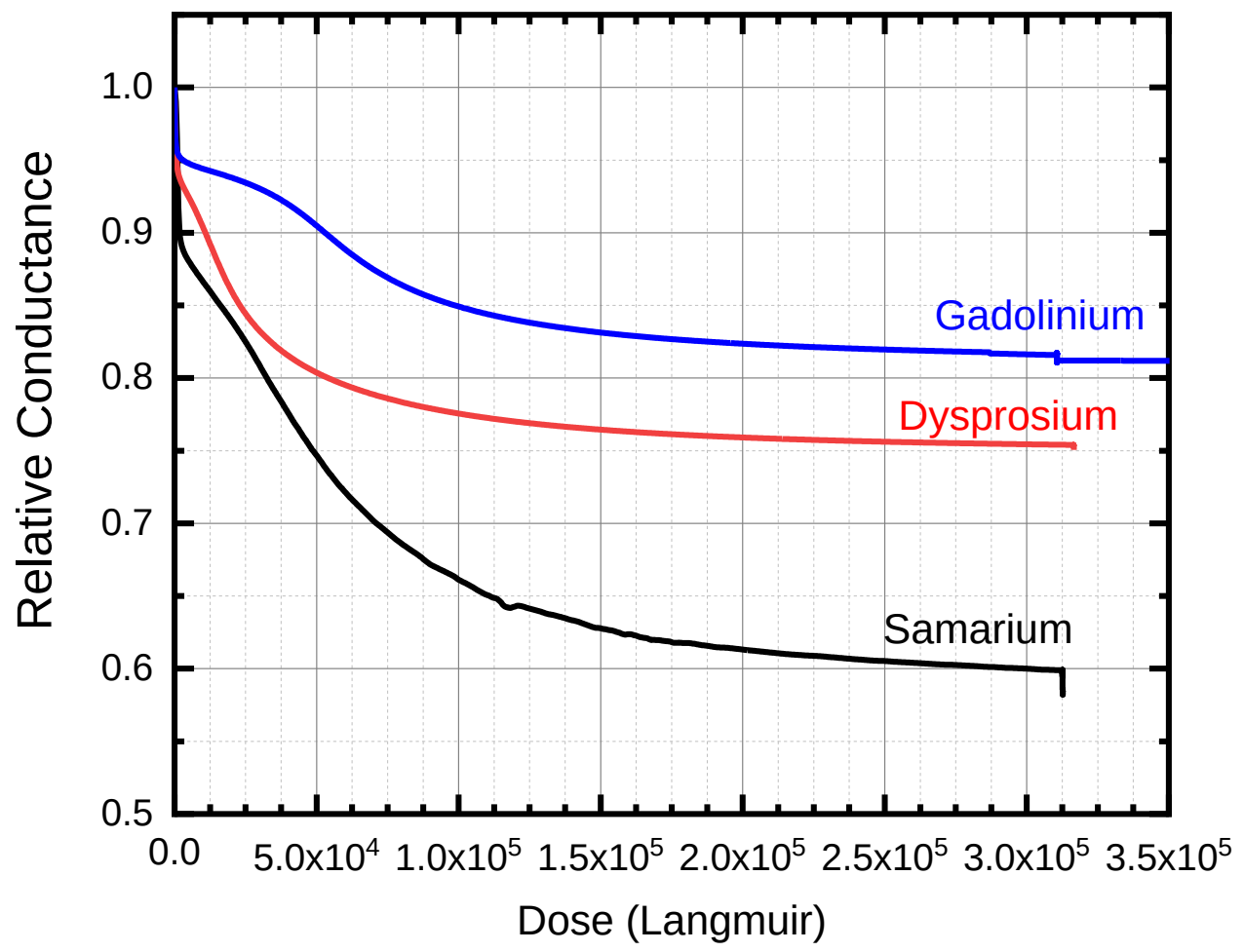

Figure 4. Relative conductance of three lanthanide thin films as they react with $N_2$.

We observe, figure 4, an initial immediate drop in conductance as soon as the film is exposed to $N_2$, then a plateau, with varying steepness and length for L elements, and finally another decrease of conductance before levelling off to a steady-state value. We attribute the initial drop to a nitridation of the top 5-10% of the lanthanide film. Remembering the >4 orders of magnitude difference in conductivity of the LN to L we can assume that layers that nitrided no longer contribute to the conductance. This difference means that the 5-10% initial drop corresponds to the top 2 nm of the lanthanide being nitrided. As more $N_2$ hits the now nitrided surface, there are fewer unbound L sites to crack the $N_2$. This necessitates a transition to either (a) atomic nitrogen diffusion from the LN to the L, or (b) molecular $N_2$ diffusing through the LN to react with L. Given the polycrystalline nature of our LN films, as seen in Figure 1, it is likely that nitrogen diffuses through grain boundaries either in atomic or molecular format. Our first attempt to model the nitridation process after the initial drop suggests a process similar to silicon oxidation based on the Deal-Grove model.[19] As more of the L turns into LN the conductance continues to drop, but the conductance change slows down further, as the diffusion through the LN layer (or the cracking efficiency at an almost fully nitrided surface) is now the rate-limiting step. After 40 minutes of $N_2$ exposure, the conductance has reached steady-state for most films.

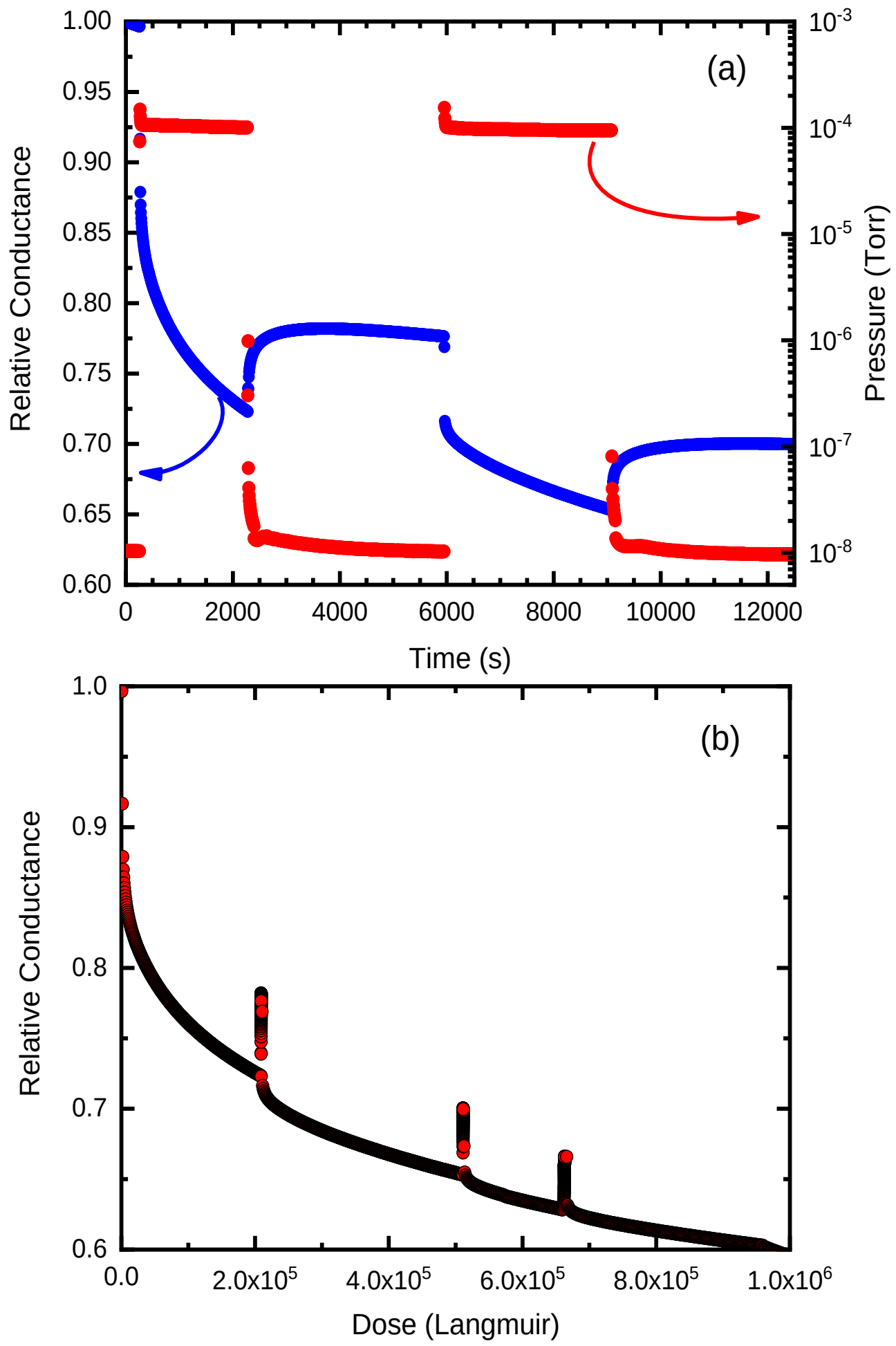

Figure 5. (a) A SmN film grown with a large $V_N$ concentration. As the film is cycled through $N_2$ and UHV conditions, the film conductivity is partially recovered during UHV conditions. (b) Same data plotted against dose for ease of comparison with (a).

Since we have described the top 5 to 10 monolayers to be more reactive, we assume these would be easier to remove nitrogen and/or $N_2$ from a LN film. To study this, we made a polycrystalline SmN thin film with a high concentration of $V_N$. The conductance of this SmN film is monitored as the sample environment is cycled between $N_2$ growth pressure ($10^{-4}$ Torr) and UHV ($10^{-8}$ Torr) at room temperature, as seen in Figure 5a. As the sample gas environment changes, so does the conductance. When $N_2$ is introduced at growth pressure levels used in typical growths in the UHV chamber the SmN conductance decreases, which indicates a filling of $V_N$ in the lattice. When the chamber is under UHV conditions the conductance of the sample increases, indicating the creation of more $V_N$. The increase of the conductance ~ 10% correlates well with the experiments described in Figure 4, indicating that the top 10% of the film is easily accessible for reactions. The creation of $V_N$ under UHV conditions could be explained by $10^{-8}$ Torr being below the equilibrium nitrogen pressure for SmN, so nitrogen is released from the SmN lattice, though no measurements of the equilibrium pressure for N in SmN exist. The relative conductance change of the SmN film is comparable to the initial nitridation of the pure Sm film, which is surprising as the SmN film is already mostly nitrided. It is also interesting to note that the SmN in these experiments is de-nitrided at room temperature, and the amount of nitrogen that can be extracted from the lattice may increase significantly if the temperature is increased.

In conclusion, we have demonstrated cracking and incorporation of molecular nitrogen in a variety of lanthanides, by successful growth of lanthanide nitrides, and conductance and XRD measurements highlighting the conversion of atomically clean surfaces lanthanides into their nitrides. Surprisingly this L+N→LN reaction can take place under mild conditions, room temperature and $N_2$ pressure as low as $10^{-5}$ Torr. It is also worth mentioning that our observations suggest that this reaction can only happen in a very clean environment, and that any surface contamination, with oxygen for example, will not allow the $N_2$ breaking. We have also observed the more polycrystalline the L layers are, the deeper the nitridation is, suggesting that grain boundaries contribute in the nitridation of lanthanides as the boundaries provide an accessible diffusion path to unreacted lanthanide. We have also provided evidence of nitrogen desorption from SmN thin films by conductance measurements in cycling nitrogen-vacuum atmosphere. We believe our results pave the way to further increase the interest on the reaction between lanthanide and $N_2$ not only from a fundamental point of view, but also for its potential in the nitrogen cracking industry.

## AUTHOR INFORMATION

### Corresponding Author

* Email: Franck.natali@vuw.ac.nz
* Email: Felicia.Ullstad@gmail.com

## ACKNOWLEDGMENT

We acknowledge funding from the Marsden Fund (Grant No. 13-VUW-1309), and the MacDiarmid Institute for Advanced Materials and Nanotechnology, funded by the New Zealand Centres of Research Excellence Fund. J. Chan thanks Viclink for financial support. The authors are grateful to Anna Garden and Stephanie Lambie (University of Otago, New Zealand) for fruitful discussions.

Authors are required to submit a graphic entry for the Table of Contents (TOC) that, in conjunction with the manuscript title, should give the reader a representative idea of one of the following: A key structure, reaction, equation, concept, or theorem, etc., that is discussed in the manuscript. Consult the journal's Instructions for Authors for TOC graphic specifications.

Insert Table of Contents artwork here

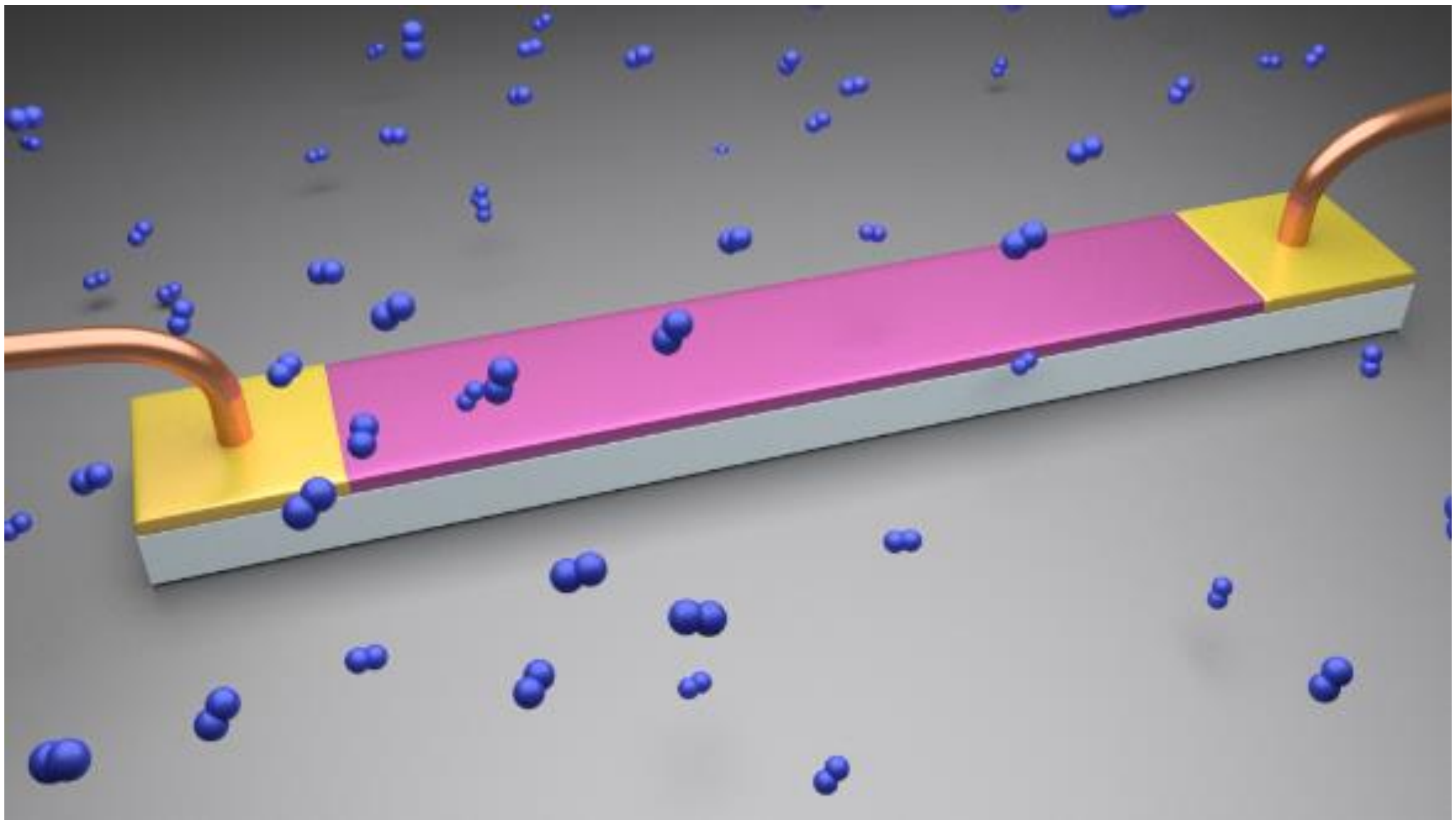